\documentclass[%
aip,
 amsmath,amssymb, 
reprint,
onecolumn 
]{revtex4-2}
\usepackage{bm}

\usepackage[utf8]{inputenc}
\usepackage[T1]{fontenc}
\usepackage{mathptmx}
\usepackage{etoolbox}

\makeatletter
\def\@email#1#2{%
 \endgroup
 \patchcmd{\titleblock@produce}
  {\frontmatter@RRAPformat}
  {\frontmatter@RRAPformat{\produce@RRAP{*#1\href{mailto:#2}{#2}}}\frontmatter@RRAPformat}
  {}{}
}%
\makeatother

\usepackage[dvipsnames]{xcolor} 
\usepackage{amsthm}
\usepackage{enumitem}
\usepackage{dsfont}
\usepackage{tikz}
\usetikzlibrary{arrows.meta}
\usepackage{hyperref}
\hypersetup{
	colorlinks=true,         
	linkcolor=MidnightBlue,          
	citecolor=MidnightBlue,
	urlcolor=MidnightBlue            
      }
      \usepackage{orcidlink}

\theoremstyle{thmstyleone}%
\newtheorem{lemm}{Lemma}
\newtheorem{thm}{Theorem}

\newcommand{\RR}{\mathds{R}} 
\newcommand{\dd}{\partial}

\DeclareMathOperator{\Ad}{Ad}
\usepackage{mathtools}
\def\dQ{{\,\mkern3mu\mathchar'26\mkern-12mu d}Q} 

\begin{document}

\preprint{JMP26-AR-00467}

\title[Heat as Gauge]{Heat as Gauge}
\author{Bryan W Roberts}
\email{b.w.roberts@lse.ac.uk}
\homepage{https://personal.lse.ac.uk/robert49/}
\affiliation{ 
Centre for Philosophy of Natural \& Social Sciences\\London School of Economics \& Political Science\\Houghton Street, London WC2A 2AE, UK
}%

\date{\today}

\begin{abstract}
We show that thermodynamics can be viewed as a gauge theory. In particular, it is a principal $\RR$-bundle over a space of observable work configurations, with gauge transformations representing changes in the `hidden' contributions to energy not captured by work. Heat defines a connection, and its curvature measures deviation from local equilibrium. Summed negative curvature on a disc is shown to give an expression of the Clausius inequality on the oriented loop bounding that disc. This allows one to derive a geometric expression of the Second Law. The framework further allows one to exhibit a thermal analogue of the Aharonov-Bohm effect. Instead of producing geometric phase shifts, this effect produces geometric entropy shifts.
\end{abstract}

\maketitle

\section{Introduction}

It is sometimes possible to shed light on one physical theory by casting it in the language of another, thus allowing conceptual tools to be shared between them. The purpose of this paper is to do so by showing how classical thermodynamics is a special case of gauge physics. We first observe that the heat-work distinction can be characterised as a fibre bundle with an Ehresmann connection. We go on to show that it can be enriched to form a principal $\RR$-bundle. Three observations follow as a result. The first makes use of two well-known facts that are rarely combined: a gauge connection has vanishing curvature if and only if it is integrable,\cite{KobayashiNomizu1963fdg,hermann1975g,nakahara1990gtp} and heat is integrable if and only if it can be locally expressed as $TdS$.\cite{caratheodory1909,bernstein1960c,boyling1968c,BryantEtAl1991} Viewing thermodynamics as a gauge theory, it follows that vanishing curvature is equivalent to the local expression of heat as $TdS$. The second observation is a formulation of a geometric expression of the Second Law. It arises as a consequence of Stokes' theorem, with the Clausius inequality expressing a condition on the summed curvature of a disc circumscribing a loop of interest. The third observation is an analogue of geometric phase in thermodynamics, of the kind found in the Aharonov-Bohm effect.\cite{aharonovbohm1959a} Instead of a shift in geometric phase, this effect gives rise to a shift in geometric entropy.

The fact that thermodynamics can be associated with an Ehresmann connection on a fibre bundle was originally proposed\cite{Roberts2026Entropy} to correct a theorem of \citeauthor{jauch1972th}.\cite{jauch1972th,jauch1975at} Jauch claimed that a certain conservation law for work implies that heat is locally equal to $TdS$. Although his argument does not succeed, the translation of his ideas into the language of fibre bundles provided a simple argument that does.\cite{Roberts2026Entropy} Here we provide a more complete mathematical development of thermodynamics as a gauge theory, showing how the Jauch entropy-temperature theorem just mentioned arises as a special case. The resulting general formulation of thermodynamics as a gauge theory developed here provides deeper insight into its geometric structure, including the possibility of geometric phase and of the Second Law. It also allows for natural connections to be drawn between the gauge formulation and other geometric approaches to thermodynamics, including the order-theoretic framework\cite{buchdahl1966t,liebyngvason1998n,liebyngvason2000fresh} and the contact geometry approach.\cite{hermann1973g,mrugala1978g,quevedo2007gt}

\section{Motivation for the Fibre Bundle Framework}\label{sec:fibre-bundle-framework}

When an insulated gas is compressed at some pressure, its change in energy can be determined entirely through the observation of its visible change in volume. This change is given locally by $-PdV$. However, when the same process is repeated with the insulation removed, an additional phenomenon can occur: the change in energy can be smaller than in the previous case. If energy is not destroyed, then one is forced to conclude that the observable contribution of $-PdV$ or `work' is incomplete. Some hidden degrees of freedom beyond $V$ must be involved. Although early nineteenth-century engineers did not yet understand their nature, they were still able to formulate the theory of thermodynamics. The strategy is just to include an additional quantity expressing the transfer of energy into hidden degrees of freedom. In thermodynamics, that quantity is called `heat'.

This thinking provides a general perspective on thermodynamics that applies to any physical system with hidden degrees of freedom. In a general mechanical system, change in work energy is given in local coordinates $(V_1,\dots,V_n)$ by $\rho := -P_1dV_1 - \cdots - P_ndV_n$, where each $P_i$ is a smooth function, and each $P_idV_i$ represents a contribution to energy. The number $n$ of these contributions is often very large. However, we take thermodynamics to apply in situations where we only have observable access to a small number of them. A typical example is a two-cylinder engine described by two volume degrees of freedom $(V_1,V_2)$. Only a portion of the total energy can generally be applied or extracted as work, $\omega = -P_1dV_1 - P_2dV_2$. The contributions to a change in energy from the remaining `hidden' degrees of freedom of the gas are what we will call heat. This is not to say that the effects of that heat are entirely unobservable. We will discuss the precise sense in which they are. Rather, the proposal here is to view thermodynamic work as the cut-off of total mechanical work, which restricts attention to just the subset of observable degrees of freedom. The remaining contributions to energy beyond some $m<n$ are then represented using a single energy parameter.\footnote{In this paper heat consists of \emph{all} non-work contributions to energy. This means that some non-work quantities that are sometimes not considered to be heat, such as energy transfer through the stirring of a fluid, will be characterised as heat for our purposes. If one does not wish to view stirring as heat, then in this framework one must identify its degrees of freedom and formulate it as a contribution to the work energy of a process.} We will show that this fundamental distinction contains the essential structure of thermodynamics, in both equilibrium and non-equilibrium forms.

Our strategy will be to express the heat-work distinction using a fibre bundle with one-dimensional fibres. We first identify its thermodynamic properties, and then illustrate them with simple examples. The observable energy of work will be defined to be non-trivial only in horizontal directions, while the hidden energy of heat is always non-trivial in vertical directions, whose coordinate value will be interpreted as entropy.

The framework begins with two smooth manifolds $P$ and $M$, with $\dim P = \dim M+1$. Let
\begin{equation}
  \pi:P\to M
\end{equation}
be a fibre bundle. One could at this stage introduce local coordinates $(V_1,\dots, V_m)$ for $M$, with adapted local coordinates $(V_1,\dots,V_m,s)$ for $P$, and then define \emph{work} to be the one form on $P$ given in some neighbourhood $B\subseteq P$ of every point by
\begin{equation}\label{eq:work}
  \omega := -\sum_{i=1}^{m}P_idV_i,
\end{equation}
where each $P_i:B\to\RR$ is a smooth function for $i=1,\dots,m$. We will instead formulate this property in an equivalent geometric form, defining work to be any one form $\omega$ on $P$ that annihilates vertical vectors,
\begin{equation}\label{eq:work-kernel}
  \ker\pi_* \subseteq \ker\omega.
\end{equation}
Such a one form is called \emph{semi-basic}. This property is equivalent to that of Equation \eqref{eq:work}. We adopt it here because it more directly expresses the idea that the vertical directions are `invisible' on the manifold $M$ of observable degrees of freedom.

The remaining contribution to the energy of a given process is \emph{heat}, which many authors denote $\dQ$. We will define heat to be a smooth one form $\xi$ on $P$ that is vertical in the sense that
\begin{equation}
  \ker\xi \cap \ker\pi_* = \{0\}.
\end{equation}
This captures the fact there is always heat in a local process described by a vertical vector $X\in\ker\pi_*$. Note that $\xi$ is nowhere-vanishing as a consequence of this definition. That is a natural restriction on the scope of thermodynamics: it still allows some curves to be adiabatic or `heat-free', but restricts attention to systems in which heat is always non-zero for some process at every point.

The total energy of a local process given by a vector $X$ on $P$ is the sum of work and heat energy, $\omega(X) + \xi(X)$. In thermodynamics we assume that this total energy is \emph{locally conserved} in the sense of being closed, $d(\omega + \xi)=0$. Equivalently, this says that in some neighbourhood $B$ of every point, the total energy produced along every closed loop $c$ in $B$ is zero, $\int_c(\xi+\omega)=0$. This statement is much weaker than the statement that energy is constant along every process, which is not required in thermodynamics. However, since every closed form is locally exact, it follows that in a neighbourhood $B$ of every point in $P$, there is a smooth function $U:B\to\RR$ such that in this neighbourhood,
\begin{equation}
  dU = \xi + \omega.
\end{equation}
This is our expression of the First Law of thermodynamics. We will refer to the local smooth function $U$ as \emph{total energy}. In summary, a minimal \emph{fibre bundle framework for thermodynamics} consists of the following structures:
\begin{itemize}
  \item a fibre bundle $\pi:P\to M$ with one-dimensional fibres, with $M$ representing work configurations;
  \item a semi-basic one form $\omega$ on $P$ representing work energy;
  \item a one form $\xi$ on $P$ that is vertical ($\ker\xi\cap\ker\pi_*=\{0\}$) and represents heat energy;
  \item a closure postulate $d(\omega+\xi)=0$, where the local function $U$ satisfying $dU=\omega+\xi$ represents total energy.
\end{itemize}

To illustrate with an example: let $M=\RR^+\times \RR^+$ be the manifold of volume configurations for a two-cylinder engine, with global coordinates $(V_1,V_2)$. The changes in state as it proceeds through its cycle are described by a closed loop $\lambda$ in $M$. However, the work energy $\omega = -P_1dV_1 - P_2dV_2$ is not necessarily the total energy produced by this process. In a realistic engine, the cycle will also involve `hidden' contributions to energy associated with a spiral in the total manifold $P$. This spiral projects down onto the closed curve $\lambda$ in $M$ shown in Figure \ref{fig:engine}.

\begin{figure}[tbh]\begin{center}
\begin{tikzpicture}[
  x={(0.52cm,0.55cm)}, 
  y={(1cm,0cm)},      
  z={(0cm,1cm)},      
  font=\small,
  >=Stealth,
  declare function={
    vone(\t)=2.5+1.1*sin(\t);
    vtwo(\t)=4.05+2.0*cos(\t)-0.572*sin(\t);
    energy(\t)=1.5+2.6*(\t+60)/360;
  }
]
  \newcommand{\curvearrow}[2]{%
    \pgfmathsetmacro{\dx}{-2*sin(#1)}%
    \pgfmathsetmacro{\dy}{0.605*cos(#1)+#2*2.6/(2*pi)}%
    \pgfmathsetmacro{\speed}{veclen(\dx,\dy)}%
    \pgfmathsetmacro{\backangle}{#1-0.16*180/(pi*\speed)}%
    \pgfmathsetmacro{\dx}{-2*sin(\backangle)}%
    \pgfmathsetmacro{\dy}{0.605*cos(\backangle)+#2*2.6/(2*pi)}%
    \pgfmathsetmacro{\speed}{veclen(\dx,\dy)}%
    \pgfmathsetmacro{\nx}{-\dy/\speed}%
    \pgfmathsetmacro{\ny}{\dx/\speed}%
    \pgfmathsetmacro{\tailx}{0.52*vone(\backangle)+vtwo(\backangle)}%
    \pgfmathsetmacro{\taily}{0.55*vone(\backangle)+#2*energy(\backangle)}%
    \pgfmathsetmacro{\tipx}{0.52*vone(#1)+vtwo(#1)}%
    \pgfmathsetmacro{\tipy}{0.55*vone(#1)+#2*energy(#1)}%
    \begin{scope}[x={(1cm,0cm)},y={(0cm,1cm)}]
      \draw[line width=0.9pt,line cap=round,line join=round]
        ({\tailx+0.085*\nx},{\taily+0.085*\ny}) --
        (\tipx,\tipy) --
        ({\tailx-0.085*\nx},{\taily-0.085*\ny});
    \end{scope}%
  }

  \draw[->,line width=0.7pt]
    (0,0,0) -- (5.5,0,0) node[above left] {$V_1$};
  \draw[->,line width=0.7pt]
    (0,0,0) -- (0,8.0,0) node[right] {$V_2$};
  \draw[->,line width=0.7pt]
    (0,0,0) -- (0,0,5) node[above] {$S$};

  \foreach \t in {0,190}{
    \draw[black!45,densely dashed,line width=0.55pt]
      ({vone(\t)},{vtwo(\t)},0) --
      ({vone(\t)},{vtwo(\t)},{energy(\t)});
  }

  \draw[black!45,densely dashed,line width=0.55pt]
    ({vone(-60)},{vtwo(-60)},0) --
    ({vone(300)},{vtwo(300)},{energy(300)});

  \draw[line width=0.9pt]
    plot[domain=-60:300,samples=181,variable=\t]
      ({vone(\t)},{vtwo(\t)},0);
  \curvearrow{90}{0}
  \curvearrow{270}{0}

  \draw[
    line width=1.1pt,
    preaction={draw=white,line width=3.3pt}
  ]
    plot[domain=-60:300,samples=241,variable=\t]
      ({vone(\t)},{vtwo(\t)},{energy(\t)});
  \curvearrow{-15}{1}
  \curvearrow{150}{1}

  \fill ({vone(-60)},{vtwo(-60)},0)
    circle[radius=1.4pt];
  \fill ({vone(-60)},{vtwo(-60)},{energy(-60)})
    circle[radius=1.6pt]
    node[below right] {}; 
  \fill ({vone(300)},{vtwo(300)},{energy(300)})
    circle[radius=1.6pt]
    node[above right,yshift=4pt] {}; 

  \draw[->,black!60,line width=0.55pt,xshift=0.28cm]
    ({vone(0)},{vtwo(0)},1.30) --
    ({vone(0)},{vtwo(0)},0.35)
    node[midway,right] {$\pi$};
\end{tikzpicture}
    \caption{Depiction of a two-cylinder engine's cycle as a spiral in a fibre bundle.}\label{fig:engine}
  \end{center}\end{figure}

The physical origin of this spiral is of course the exchange of energy involving hidden degrees of freedom, such as the microscopic molecular motion of the gas and its container. However, the fibre bundle framework is much more general than the theory of gases. It can be applied to any system in which there are hidden degrees of freedom. The structures just defined are already powerful enough to draw interesting conclusions about equilibrium, entropy, and temperature. These conclusions will make use of the theory of curved connections. The spiral diagram of Figure \ref{fig:engine} is already a hint that thermodynamics contains a natural notion of holonomy.\cite{Roberts2026Entropy} In the next two sections we will identify what can already be concluded using the concept of an Ehresmann connection, before turning to the richer results available from the theory of principal connections.

\section{Heat Defines an Ehresmann Connection}\label{sec:ehresmann-connection}

Much of the content of thermodynamics is known to reside in the structure of \emph{adiabats}, or curves $\gamma$ along which heat vanishes, in that $\xi(\bar{\gamma})=0$ for each tangent vector $\bar{\gamma}$ to $\gamma$. However, the mathematical structure used to express this is usually an `adiabatic accessibility' order relation.\cite{caratheodory1909,LiebYngvason1999} Here we will make use of no such order relation. Instead, we focus only on the directions in which adiabats `point', namely the vectors in $H=\ker\xi$. In particular, we will follow a recent proposal\cite{Roberts2026Entropy} in observing that the distribution $H$ defines an Ehresmann connection with a notion of curvature. In the next section, we will show that this connection is flat if and only if $\xi=TdS$ in a neighbourhood of every point.

A smooth subbundle $H \subseteq TP$ is called an \emph{Ehresmann connection} for the fibre bundle $\pi:P\to M$ if it is a horizontal distribution,
\begin{equation}
  T_pP = H_p \oplus V_p,
\end{equation}
for each $p\in P$, where $V_p = \ker\pi_*|_p$ is the subspace of vertical vectors at $p$. A vector $X\in H$ is called \emph{horizontal}. This is a standard generalisation\cite{KobayashiNomizu1963fdg,hermann1975g} of the notion of a principal connection on a principal $G$-bundle. It follows almost immediately from our definitions that the heat one form $\xi$ for thermodynamics defines an Ehresmann connection $H=\ker\xi$, with horizontal vectors $X\in H$ arising as tangent vectors to the adiabats. We verify this in the following.

\begin{lemm}\label{lemm:1}
  Let $\pi:P\rightarrow M$ be a fibre bundle with one-dimensional fibres. Let $\xi$ be a smooth one form on $P$ satisfying $\ker\pi_*\cap\ker\xi = \{0\}$. Then $H=\ker\xi$ is an Ehresmann connection.
\end{lemm}
\begin{proof}
Since $H=\ker\xi$ and $\xi$ is nowhere-vanishing, $\dim H_p = \dim P-1 = \dim M$. So, since $\pi:P\to M$ has one-dimensional fibres, $\dim V_p = \dim\ker\pi_*|_p=1$, and we get the required decomposition $T_pP = H_p \oplus V_p$ for each $p\in P$.
\end{proof}

Like a principal connection, which we will turn to shortly, an Ehresmann connection $H$ defines a notion of \emph{curvature}.\cite{hermann1975g} This is the map $\Omega_H$ on horizontal vector fields given by
\begin{equation}\label{eq:ehresmann-curvature}
  \Omega_H(X,Y) := -\mathrm{vert}[X,Y].
\end{equation}
Here, `$\mathrm{vert}$' denotes the projection onto the vertical subspace $V$, and the bracket is the Lie bracket. We will shortly show that it coincides with the curvature of a principal connection, once our fibre bundle is enriched to include the structure of a principal $G$-bundle with a compatible principal connection. However, an Ehresmann connection alone allows one to draw substantial conclusions about thermodynamics. We begin with one such conclusion.

\section{Zero Curvature as Equilibrium}\label{sec:equilibrium}

Equilibrium has many definitions in thermodynamics. Here we will consider it to be the property that, in a neighbourhood of every point, heat can be written as $\xi = TdS$ for some smooth functions $T$ and $S$. This definition of `local equilibrium' will shortly be motivated through considerations of the Second Law. In this section, we will take it for granted. In the fibre bundle framework with a heat one form $\xi$, one has the following result: local equilibrium in the sense of the local expression $\xi=TdS$ is equivalent to the vanishing of curvature or `flatness' of the connection $H=\ker\xi$.  The proof of this fact is a simple combination of two classical results: zero curvature implies integrability, and integrability implies $\xi=TdS$.

A smooth rank-$k$ distribution $H\subseteq TP$ on a manifold $P$ of dimension $n$ is called \emph{integrable} if each point $p\in P$ admits an open neighbourhood $B$ and a smooth submersion $\varphi:B\to \RR^{n-k}$ whose level sets are \emph{integral submanifolds} each with its tangent space at every point equal to $H_p$. By the celebrated Frobenius theorem, a smooth distribution $H$ is integrable if and only if it is \emph{involutive}, $[X,Y]\in H$ for all $X,Y\in H$. It is also a cornerstone result of gauge theory that a distribution $H$ is involutive if and only if its Ehresmann curvature $\Omega_H$ vanishes (Theorem 5.2 of \citeauthor{KobayashiNomizu1963fdg}\cite{KobayashiNomizu1963fdg} and Theorem 2.1 of \citeauthor{hermann1975g}\cite{hermann1975g}). In summary, we have the following.
\begin{lemm}\label{lemm:2}
  An Ehresmann connection $H=\ker\xi$ on a fibre bundle has vanishing curvature if and only if it is integrable.
\end{lemm}

This is the first classical result. The second\cite{boyling1968c,BryantEtAl1991} is that the integrability of $H=\ker\xi$ is equivalent to the local existence of functions $T$ and $S$ such that $\xi=TdS$. In thermodynamics, these functions are generally interpreted as \emph{temperature} and \emph{entropy}, although we will not motivate that interpretation until later sections. We record this result as:

\begin{lemm}\label{lemm:3}
An Ehresmann connection $H = \ker\xi$ on a fibre bundle with $\xi$ nowhere-vanishing is integrable if and only if, in a neighbourhood of every point, there exist smooth functions $T,S$ such that $\xi=TdS$.
\end{lemm}
\begin{proof}
Integrability is equivalent to the existence of a local foliation by integral submanifolds, with the level sets of a smooth local submersion defining the leaves of the foliation. Thus, $\ker\xi$ is integrable only if there is a local function $S$ such that $\ker\xi = \ker dS$. This implies that $\xi$ and $dS$ are proportional, $\xi=TdS$ for some smooth non-vanishing function $T$. Conversely, if $\xi=TdS$ in a neighbourhood of every point, then since $\xi$ is nowhere-vanishing, $T$ and $dS$ are nowhere-vanishing as well, and so $H=\ker dS$. The level sets of $S$ then provide the local integral submanifolds that ensure $H$ is integrable.
\end{proof}

Combining all three of our lemmas we now have:
\begin{thm}\label{thm}
Let $\pi:P\rightarrow M$ be a fibre bundle with one-dimensional fibres. Let $\xi$ be a smooth one form on $P$ satisfying $\ker\pi_*\cap\ker\xi = \{0\}$. Then $H=\ker\xi$ is an Ehresmann connection, and the following are equivalent: 
  \begin{enumerate}
    \item[$(A)$] in a neighbourhood of every point, there exist smooth functions $T,S$ such that $\xi = TdS$.
    \item[$(B)$] $H$ has vanishing curvature.
  \end{enumerate}
\end{thm}

One way to interpret this is the following. In ordinary gauge theory, the curvature of an Ehresmann connection measures local field strength, or deviation of the system from one of vanishing local field strength. In thermodynamics, the curvature of $H=\ker\xi$ measures the deviation of the system from local equilibrium. Thus, when curvature vanishes, we get one of the defining local relations of equilibrium thermodynamics, $\xi=TdS$, as a result of Theorem \ref{thm}.

It has been pointed out\cite{Roberts2026Entropy} that, by viewing local equilibrium as the vanishing of the curvature of an Ehresmann connection, $\xi=TdS$ follows locally from a conservation law due to \citeauthor{jauch1972th}.\cite{jauch1972th,jauch1975at} Suppose that an adiabatic curve $\gamma$ in $P$ projects down onto a closed curve $\lambda:=\pi\circ\gamma$ in the work configuration space, as it does in the cycle of a two-cylinder engine. Then vanishing curvature follows whenever, in a neighbourhood of every point, every horizontal lift $\gamma$ of a sufficiently small loop $\lambda$ is closed, which is to say that every local holonomy is trivial. By the First Law $dU=\xi+\omega$, this latter statement implies that no work is produced during the adiabatic process $\gamma$,
\begin{equation}
  \int_\gamma\omega = 0,
\end{equation}
since $\int_\gamma dU=0$ when $\gamma$ is closed and $\int_\gamma\xi=0$ when it is an adiabat. Jauch's statement can be viewed as a `reversible version' of Kelvin's law, that in a local neighbourhood of each point, a process that is a cycle in the work variables and involves no heat exchange cannot perform work. Since it follows from this that curvature vanishes, Jauch's assumption can thus be used as a basis for the claim that local functions exist such that $\xi=TdS$ in equilibrium thermodynamics, as an alternative to Carath\'eodory's theorem.\cite{Roberts2026Entropy}

We have expressed curvature as the failure of the flows along two vectors $X,Y$ to commute in the vertical direction. Since gauge theory encodes a great deal more structure in those vertical fibres through the use of a principal $G$-bundle, it will be informative to analyse thermodynamics from that perspective as well. This is the topic of the next section; in Section \ref{sec:second-law}, we will show that it leads directly to an expression of the Second Law, and in Section \ref{sec:geometric-phase} to an analogue of geometric phase.

\section{Principal Bundle and Temperature}\label{sec:principal-bundle}

We will now cast thermodynamics in a more standard framework for Yang--Mills gauge theory. In this framework, the fibre bundle is a principal $G$-bundle $\pi:P\to M$ with a principal connection $\alpha$. The relationship between this framework and the Ehresmann one of the previous two sections is straightforward:\cite{sternberg1964dg,KobayashiNomizu1963fdg,bleecker1981g} if $\alpha$ is a principal connection, then $H=\ker\alpha$ is an Ehresmann connection; conversely, if $H$ is an Ehresmann connection that is invariant under the action of a structure group $G$, there exists a principal connection $\alpha$ such that $H=\ker\alpha$.

We begin by identifying and interpreting the structure group $G$ for thermodynamics. There are only two options, $U(1)$ and $(\RR,+)$. This is because the fibre bundle for thermodynamics has connected one-dimensional fibres: the structure group $G$ is diffeomorphic to each fibre in a principal $G$-bundle, and $U(1)$ and $(\RR,+)$ are the only connected one-dimensional Lie groups. In electromagnetism, the group is $G=U(1)$. This is appropriate for the description of phase, which is periodic. In thermodynamics, it is 
\begin{equation}
  G=(\RR,+).
\end{equation}
This is appropriate for representing the energy of hidden degrees of freedom, since it is \emph{not} periodic. Principal $\RR$-bundles of this kind are standard examples in differential geometry. They rarely appear in Yang--Mills gauge theories, the structure groups of which are usually compact. However, our considerations show that their natural role is in the study of thermodynamics.

The action of the structure group does not represent `pure descriptive redundancy' in thermodynamics. This is because the base space $M$ does not represent spacetime. Instead, $M$ is the manifold of work configurations, which are by definition the system's observable degrees of freedom. As a result, the action of the structure group represents changes in energy that make no \emph{observable} difference, rather than those that make no difference at all. Such changes may still represent a physical difference to a more complete underlying description. Note that our use of `observable' here refers specifically to the work configuration variables $V_i$ on $M$. The work one form $\omega$ may still depend on the hidden state, and so need not be invariant under fibre translations. We will discuss an example in which this is the case in Section \ref{sec:geometric-phase}.

Interpreting the structure group in this way imposes an additional constraint on the heat one form $\xi$. Vertical translations are changes in the energy of hidden degrees of freedom. On this interpretation, they should not change whether a process is adiabatic, since adiabatic processes do not involve any transfer of hidden energy. An appropriate compatibility requirement to express this is that the adiabatic directions $X\in\ker\xi$ remain adiabatic under the action of the structure group. Writing that action as $R_s:P\to P$, this requirement can be expressed as
\begin{equation}
  \ker(R_s^*\xi) = \ker\xi = H \text{ for all } s\in\RR.
\end{equation}
Thus, when the fibre bundle $\pi:P\to M$ is taken to be a $G$-principal bundle, we require that the heat one form $\xi$ is not just vertical $(\ker\xi\cap\ker\pi_*=\{0\})$, but that the set of adiabats is invariant under fibre translations. This is exactly the standard mathematical definition of a \emph{principal connection} $H$: it is an Ehresmann connection that is invariant under the action of the structure group $G$.

This definition of a principal connection is well-known\cite{KobayashiNomizu1963fdg,bleecker1981g} to be equivalent to a more standard definition in physics, as a Lie algebra-valued one form $\alpha$ with the properties that (a) the fundamental field $a^*$ generated by each element of the Lie algebra $a\in\mathfrak{g}$ satisfies $\alpha(a^*)=a$, and (b) the action $R_g$ of $G$ on $P$ satisfies $R_g^*\alpha=\Ad_{g^{-1}}\alpha$. In our case of a principal $\RR$-bundle, we can replace $\alpha$ with an ordinary one form on $P$. Condition $(b)$ then simplifies to $R_r^*\alpha=\alpha$ due to the fact that $(\RR,+)$ is abelian, where $R_r$ is the action defined on local coordinates $(S,V)$ by $R_r(S,V)=(S+r,V)$ for all $r\in\RR$. This one form satisfies $\ker\alpha = H = \ker\xi$. Since both are non-vanishing, it follows that $\xi$ and $\alpha$ are proportional, in that there exists a smooth nowhere-vanishing function $T:P\to\RR$ such that $\xi = T\alpha$. We thus have:

\begin{lemm}\label{lemm:connection}
  Given a principal $\RR$-bundle $\pi:P\to M$ with a smooth one form $\xi$, the following are equivalent.
  \begin{enumerate}
    \item[(1)] $\xi$ satisfies the requirements of a heat one form, $\ker\xi\cap\ker\pi_*=\{0\}$ and $\ker (R_r^*\xi) = \ker\xi$ for all $r\in\RR$.
    \item[(2)] There exists a smooth nowhere-vanishing function $T:P\to\RR$ such that $\alpha=\xi/T$ is a principal connection.
  \end{enumerate}
\end{lemm}

This function $T$ is a candidate for physical \emph{temperature} in the following sense. Using the First Law $dU = \omega + \xi$, together with the fact that work $\omega$ annihilates vertical vectors, we have that if $S$ is a principal coordinate function satisfying $S(R_r(p))=S(p)+r$, then
\begin{equation}
  \frac{\dd U}{\dd S} = dU(\partial_S) = \xi(\partial_S).
\end{equation}
So, since $\xi=T\alpha$ and $\alpha(\dd_S)=1$ for a principal connection, $\dd U/\dd S = T\alpha(\dd_S) = T$. Thus, $T$ can represent temperature in the sense of a change in energy in the vertical direction, corresponding to hidden contributions to energy.

Theorem \ref{thm} implies that when $H=\ker\xi=\ker\alpha$ is a flat Ehresmann connection, every point admits a neighbourhood $B$ with functions $T_B$ and $S_B$ such that $\xi=T_BdS_B$. But, when $\alpha=\xi/T$ is a flat \emph{principal} connection, the Poincar\'e lemma provides a simpler route to this result. Namely, the closure $d\alpha=0$ of a flat connection implies there is a local function $S_B$ such that $\xi/T=dS_B$. Thus, when $\alpha$ is flat, one can always find appropriate $T_B$ and $S_B$ such that the $T_B$ of Theorem \ref{thm} is equal to $T$ in the definition of $\alpha$ throughout the neighbourhood. Thus, these two notions of a `temperature function' align in the flat case, but the function $T$ in $\alpha=\xi/T$ is more general because it is defined even when $\alpha$ is curved. In the next section we will see that a similar alignment holds between $S_B$ and the fibre coordinate function $S$, up to a gauge transformation.

In summary, two further structures allow one to view thermodynamics not just as a fibre bundle theory, but as a theory on a principal $\RR$-bundle:
\begin{itemize}
  \item an abelian structure group $G=(\RR,+)$ that preserves adiabats in the sense of leaving the Ehresmann connection $H=\ker\xi$ invariant; and
  \item a principal connection $\alpha=\xi/T$ associated with a smooth nowhere-vanishing function $T:P\to\RR$, arising from the assumption that adiabats remain adiabats under fibre translations.
\end{itemize}

\section{Curvature and Gauge Invariance}

A principal connection, like an Ehresmann connection, admits a notion of curvature. These two notions align in a simple way for the present case of an abelian principal $G$-bundle. The reason is, if $\mathcal{F} = d\alpha$ is the curvature of the principal connection, and $\dd_S$ is the canonical vertical vector field for some principal fibre coordinate $S$, then a standard calculation\cite{sternberg1964dg,KobayashiNomizu1963fdg,bleecker1981g} shows that
\begin{equation}
  \mathcal{F}(X,Y)\dd_S = \Omega_H(X,Y)
\end{equation}
for horizontal vectors $X,Y$. In particular, Ehresmann curvature $\Omega_H$ vanishes in horizontal directions if and only if principal curvature $\mathcal{F}$ does.

In a local trivialisation\footnote{In a principal $\RR$-bundle, unlike a principal $U(1)$-bundle, a global trivialisation always exists. We still often refer to a local section here for the purpose of describing certain local facts.} $\Sigma\times\RR$ with $\Sigma\subseteq M$, these structures can be expressed in a familiar form as fields on the manifold $M$ with respect to a given section $\sigma:\Sigma\to P$,
\begin{equation}\label{eq:fields-on-M}
  A := \sigma^*\alpha, \qquad F := \sigma^*\mathcal{F} = dA.
\end{equation} 
We say that a field on $M$ is \emph{gauge invariant} if and only if its expression is unchanged by each local gauge transformation, represented by a smooth function $\chi:\Sigma\to\RR$ such that $\sigma(V)\mapsto \sigma_\chi(V) := R_{\chi(V)}\left(\sigma(V)\right)$, where $R_r$ is the action that transforms local coordinate functions $(S,V)$ as $R_r(S,V)=(S+r,V)$. Writing $S_\sigma(V) := S(\sigma(V))$, we find that $R_{\chi}$ corresponds to the local gauge transformation,
\begin{equation}
  S_\sigma \mapsto S_\sigma + \chi, \qquad A \mapsto A + d\chi.
\end{equation}

In these statements, $S$ is a vertical coordinate function. For a flat connection $\alpha=\xi/T$, we also have a function $S_B$ in a neighbourhood $B$  of every point such that $\xi=TdS_B$. But, these two notions of an `entropy function' $S_B$ and $S$ align in $B$ up to a local gauge transformation: $\alpha=\xi/T=dS_B$ and $\alpha(\dd_S)=1$ together imply that $\partial_s(S_B-S)=dS_B(\dd_S)-\dd_S(S)=\alpha(\dd_S)-1=0$. Thus, for some smooth function $\chi$, we have that at every point $p\in B$ written in local coordinates as $(S,V)$,
\begin{equation}
  S_B(S,V)=S+\chi(V).
\end{equation}

As usual in an abelian Yang--Mills theory, the curvature field $F=dA$ is gauge invariant, since $F\mapsto d(A+d\chi) = dA = F$, and the potential field $A$ is not. However, as remarked above, local gauge transformations are not descriptive redundancy in thermodynamics. They are rather changes that are unobservable from the perspective of the manifold of work configurations $M$. As a result, gauge invariance is not a requirement for physical significance in thermodynamics.

That said, gauge invariance still plays an important physical role. Gauge invariance in thermodynamics is a necessary condition for a quantity to be observable up to the limitations of one's measurement abilities. When observations are restricted to a manifold $M$ of work configurations, like the volumes of an engine's cylinders, those observations are all insensitive to differences related by a gauge transformation, since the vertical fibre translations transform the energy of degrees of freedom that are hidden. Thus, the vertical coordinate function $S$ is not observable because it is not gauge invariant. But the difference $S(p)-S(q)$ between two points $p,q$ on the same vertical fibre \emph{is} gauge invariant. In the next section, we will show that its interpretation as a change in entropy is associated with a version of the Second Law that is controlled by curvature.

\section{The Second Law}\label{sec:second-law}

The principal bundle framework for thermodynamics provides a natural expression of the law of non-decreasing entropy. It is exactly analogous to Clausius' expression of it: given the Clausius inequality $\int_c \xi/T\leq 0$ on closed curves, adiabatic cycles have non-decreasing entropy. The generality and rigour of Clausius' argument have been challenged by Carath\'eodory and others.\cite{caratheodory1909,uffink-bluffyourway} However, we will show that the principal bundle framework provides a simple and rigorous expression of it. We then show that the inequality derives from a property of curvature. From this perspective, the Second Law follows from a geometric expression of the Clausius inequality together with an application of Stokes' theorem. We use this to help to motivate the interpretation of the vertical coordinate as entropy and vanishing curvature as local equilibrium.

Let $\pi:P\to M$ be a principal $\RR$-bundle with a principal connection $\alpha=\xi/T$. An adiabat $\gamma:[0,1]\to P$ satisfies $\xi(\bar{\gamma}) = 0$, and so $\alpha(\bar{\gamma})=0$ as well. Suppose that the adiabat $\gamma$ projects down onto a closed curve $\lambda:=\pi\circ\gamma$ in $M$, so that the system describes a cycle of the kind depicted in Figure \ref{fig:engine}. Now connect the vertically separated endpoints of $\gamma$ by a vertical curve $\nu$ from $\gamma(1)$ to $\gamma(0)$. For vertical curves $\nu$ and a vertical coordinate function $S$ satisfying $S(R_r(p))=S(p)+r$, a principal connection satisfies $\alpha(\bar{\nu})=dS(\bar{\nu})$. Thus, writing $S_1:=S(\gamma(0))$ and $S_2:=S(\gamma(1))$, we have that $\int_\nu\alpha = S_1 - S_2$. Combining this with the fact that $\int_\gamma\alpha=0$, we now have that for the completed loop $c$,
\begin{equation}\label{eq:pre-inequality}
  \int_c \frac{\xi}{T} = \int_c\alpha = \underbrace{\int_\gamma\alpha}_{0} + \int_\nu\alpha = \int_\nu dS = S_1-S_2.
\end{equation}
If the closed curve $c$ in $P$ satisfies the Clausius inequality, $\int_c\xi/T\leq0$, then Equation \eqref{eq:pre-inequality} implies
\begin{equation}
  S_1 - S_2 = \int_c \frac{\xi}{T} \leq 0.
\end{equation}
Thus, the change in entropy after the process $\gamma$ is complete is non-negative, $S_1 \leq S_2$. This is the Clausius law of non-decreasing entropy, provided that the shifts in hidden energy along a vertical coordinate are interpreted as changes in entropy. We summarise this law in the following.

\begin{thm}\label{thm:2}
 Let $\pi:P\to M$ be a principal $\RR$-bundle with a principal connection $\alpha=\xi/T$ and a fibre coordinate $S$ such that $S(R_r(p))=S(p)+r$. Let $\gamma:[0,1]\to P$ be an adiabat ($\xi(\bar\gamma)=0$) whose projection onto $M$ is closed, and write $S_1:=S(\gamma(0))$ and $S_2:=S(\gamma(1))$. Let $\nu$ be a vertical curve from $\gamma(1)$ to $\gamma(0)$, with $c$ denoting the completed loop. Then
\begin{equation}
  S_2-S_1 =-\int_c\frac{\xi}{T}.
\end{equation}
If the loop $c$ satisfies the Clausius inequality
\begin{equation}
  \int_c\frac{\xi}{T}\leq0,
\end{equation}
then $S_1 \leq S_2$, with strict inequality if and only if the Clausius inequality is strict.
\end{thm}

The Clausius inequality contains a notion of orientation that can be elegantly expressed in terms of the curvature $\mathcal{F}=d\alpha$ of the connection. Consider the respective potential and curvature fields $A=\sigma^*\alpha$ and $F:=\sigma^*\mathcal{F}$ on $M$, defined for a given local section $\sigma$. As discussed in the previous section, $F$ is gauge invariant and independent of the choice of $\sigma$. Suppose the closed curve $\lambda=\dd\Sigma$ is the boundary of a smooth disc $\Sigma\subseteq M$, with the latter oriented so that its induced boundary orientation agrees with the direction of the tangent field to $\lambda$. Then we have by Stokes' theorem that
\begin{equation}\label{eq:stokes}
  \int_{c}\frac{\xi}{T} = \int_{c}\alpha = \int_{\dd\Sigma}A = \int_{\Sigma} dA = \int_\Sigma F,
\end{equation}
and so as a consequence,
\begin{equation}
  \int_{c}\frac{\xi}{T}\leq 0 \qquad \text{if and only if} \qquad \int_\Sigma F\leq0.
\end{equation}
This suggests that negative summed curvature can be used to capture the essential structure of the Second Law in a geometric way. This is indeed the case, as we confirm in the following.

The key step is to use the vertical principal coordinate function $S$ and the connection $\alpha$ to define
\begin{equation}
  \beta := dS - \alpha.
\end{equation}
Then $d\beta = -d\alpha = -\mathcal{F}$. The quantity $\beta$ is \emph{basic}, in that it is invariant under fibre translations, $R_r^*\beta = R_r^*dS - R_r^*\alpha = dS-\alpha = \beta$, and also horizontal, $\beta(\dd_S) = dS(\dd_S) - \alpha(\dd_S) = 1-1  = 0$. As a result, its representative $B:=\sigma^*\beta$ in a section $\sigma$ is gauge invariant. This $B$ is indeed a well-known dressed quantity associated with the unitary gauge and the Higgs mechanism.\cite{higgs1966s,Struyve2011} Its gauge invariance can be verified explicitly using $A:=\sigma^*\alpha$ and $S_\sigma:=S\circ\sigma$, so that $B=dS_\sigma - A$. Then a gauge transformation of $B$ takes the form
\begin{equation}
  B\mapsto d(S_\sigma+\chi) - (A + d\chi) = dS_\sigma - A = B.
\end{equation}

We will use this quantity to facilitate the geometric expression of the Second Law. Consider an adiabatic curve $\gamma$ projecting down onto a closed loop $\lambda=\dd\Sigma$, which circumscribes a smooth disc $\Sigma$ in $M$. Write $S_1:=S(\gamma(0))$ and $S_2:=S(\gamma(1))$ for its initial and final entropies, respectively. Then since $\alpha(\bar{\gamma})=0$ on adiabats, and since $\int_\gamma\beta = \int_\lambda B$ when $\beta$ is basic,
\begin{equation}\label{eq:entropy-curvature}
 S_2 - S_1 = \int_{\gamma}dS = \int_\gamma(\beta+\alpha) = \int_\gamma\beta = \int_{\lambda}B = \int_\Sigma dB = -\int_\Sigma F.
\end{equation}
Thus, the curvature of a disc circumscribed by an adiabatic loop in work configuration space measures the change in entropy after the cycle is complete. The \emph{geometric Clausius inequality} can now be defined as the condition that an adiabatic curve $\gamma$ in $P$ projects down onto a loop $\lambda = \dd\Sigma$ in $M$ such that the summed curvature over the disc $\Sigma$ is non-positive,
\begin{equation}
  \int_\Sigma F\leq 0.
\end{equation}
When the restriction of $F$ to the disc $\Sigma$ is nowhere-vanishing, this inequality requires $\lambda$ to run opposite to the boundary orientation induced by $F$. So, identifying `admissible' or `physically reasonable' loops with those that are oriented in this way guarantees by Equation \eqref{eq:entropy-curvature} that the initial entropy of the process is less than or equal to its final entropy, $S_1\leq S_2$. In this sense, the geometric Clausius inequality gives rise to a law of non-decreasing entropy. We summarise this result in the following.

\begin{thm}
   Let $\pi:P\to M$ be a principal $\RR$-bundle with a principal connection $\alpha=\xi/T$ and a fibre coordinate $S$ such that $S(R_r(p))=S(p)+r$. Let $F=\sigma^*(\mathcal{F})$ be the gauge invariant curvature of $\mathcal{F}=d\alpha$ for a given section $\sigma$. Let $\gamma:[0,1]\to P$ be an adiabat ($\xi(\bar\gamma)=0$) whose projection onto $M$ is the closed loop $\lambda=\dd\Sigma$, and write $S_1:=S(\gamma(0))$ and $S_2:=S(\gamma(1))$. Then
\begin{equation}
  S_2-S_1 =-\int_\Sigma F.
\end{equation}
If the loop $\lambda$ satisfies the geometric Clausius inequality
\begin{equation}
  \int_\Sigma F\leq0,
\end{equation}
then $S_1 \leq S_2$, with strict inequality if and only if the geometric Clausius inequality is strict.
\end{thm}

This geometric expression of the Second Law shows that, although the absolute value of entropy is not observable in that it is not gauge invariant as a field on $M$, its change can still be measured using the gauge invariant curvature field $F$. The law of non-decreasing entropy along adiabats then arises when the summed curvature is negative. Zero curvature then corresponds as expected to the equilibrium case in which entropy does not change. 

\section{Geometric Phase in Thermodynamics}\label{sec:geometric-phase}

We mention one final consequence of viewing thermodynamics as a gauge theory. That is the prediction of a phenomenon analogous to geometric phase. Thermal analogues of geometric phase have mainly been discussed in statistical extensions of thermodynamics.\cite{RenHanggiLi2010,SagawaHayakawa2011,Polettini2012} Conversely, analogues of thermal entropy and temperature have been proposed in ordinary models of Berry phase.\cite{grigorenko1990gp} Here we will show that geometric phase arises directly out of the geometry of thermodynamics. The physical significance of the result is a shift in entropy, rather than a shift in relative phase. One might refer to it as \emph{geometric entropy}.

Geometric phase in ordinary Yang--Mills theory can arise from non-trivial cohomology structure, most famously in the Aharonov-Bohm effect.\cite{aharonovbohm1959a,ShapereWilczek1989} Let $\pi:P\to M$ be a principal $U(1)$-bundle with a connection $\alpha$. Let $-iqA=\sigma^*\alpha$ denote the electromagnetic potential for a given local section $\sigma:\Sigma\to P$. In a region of $M$ with a non-trivial cohomology group, such as when a point is deleted from a plane representing the presence of an impenetrable solenoid, there is a gauge invariant quantity giving the degree of interference due to a phase shift around a loop $\lambda$ in $M$. In the Aharonov-Bohm effect, it is given (for $\hbar=c=1$) by
\begin{equation}
  \theta_\lambda := \exp\left(iq\int_\lambda A\right).
\end{equation}
This phase factor can take different values even when the curvature or `field strength' $F=dA$ does not. This includes the case that $F$ remains zero throughout the region. The density of a charged matter field like the Schr\"odinger wavefunction is then used to express the detectable change in interference.

A similar effect appears in equilibrium thermodynamics when the work configuration space $M$ has a non-trivial cohomology group. The simplest case is when $M=S^1$ is the circle. A physical implementation is given by a triangular rotor spinning inside a circular housing, similar to a Wankel engine (Figure \ref{fig:wankel}). Its configuration is given by the angle $\phi\in[0,2\pi)$ with respect to some axis.

\begin{figure}[tbh]\begin{center}
\resizebox{0.25\textwidth}{!}{
  \begin{tikzpicture}[line width=0.6pt]
    \draw (0,0) circle (2.18);             
    \fill (0,0) circle (2);                
    \filldraw[fill=white,line join=round]
      (60:2) -- (180:2) -- (300:2) -- cycle; 

    \draw (0,0) -- (0,1.05);               
    \draw[dashed] (0,0) -- (60:1.98);       
    \draw (60:0.55) arc (60:90:0.55);
    \node[font=\large] at (75:0.85) {$\phi$};
  \end{tikzpicture}
}
    \caption{Configuration angle $\phi$ for the triangular rotor of an engine.}\label{fig:wankel}
  \end{center}\end{figure}

Let $\pi:P\to M$ be a principal $\RR$-bundle with $M=S^1$ and a principal connection $\alpha=\xi/T$. Since $\dim M=1$, curvature automatically vanishes. Given a principal fibre coordinate function $S$, define the basic quantity $\beta:=dS-\alpha$ as in the previous section, so that $B:=\sigma^*\beta$ is gauge invariant for any section $\sigma$. Let $\gamma:[0,1]\to P$ be any adiabat, $\xi(\bar{\gamma})=0$, whose projection $\lambda:=\pi\circ\gamma$ is a closed loop. Let $S_1=S(\gamma(0))$ and $S_2=S(\gamma(1))$. Then we can define a quantity $\mathcal{E}_\lambda := \int_\lambda B$ called \emph{geometric entropy} which, given that $\alpha$ vanishes on adiabats, has the property that
\begin{equation}\label{eq:geometric-entropy-production}
   \mathcal{E}_\lambda := \int_\lambda B = \int_\gamma\beta = \int_\gamma dS = S_2-S_1.
\end{equation}

This is the thermodynamic analogue of geometric phase. It is a gauge invariant property of an adiabatic cycle in work configuration space. Remarkably, even though the description is in local equilibrium in the sense of having zero curvature, the quantity $\mathcal{E}_\lambda$ can still measure a non-trivial change in entropy. The curvature formulation of the Second Law does not explain this. For, as we have seen in the previous section, that formulation assumes the loop $\lambda$ in configuration space is bounded by a smooth disc. No such disc exists when the loop is non-contractible, and so that formulation does not apply.

Like the Aharonov-Bohm effect, this effect is `geometric' in that geometric entropy vanishes whenever the cohomology group $H^1_{dR}(M;\RR)$ of the base space is trivial. This is because, if $M$ were simply connected, the quantity $B$ could be written as $B=df$ for some globally defined smooth function $f:M\to\RR$. Then since $\lambda$ is a closed loop we would have that
\begin{equation}
  \mathcal{E}_\lambda = \int_\lambda B = \int_\lambda df = 0. 
\end{equation}
Thus, on a simply connected base space, geometric entropy does not arise.

The situation changes when the work configuration space is $M=S^1$. Then there need not be any smooth function $f$ such that $B=df$, in which case $\mathcal{E}_\lambda$ can be non-zero. This effect is indeed even measurable. However, unlike in the Aharonov-Bohm effect, the measurement is not expressed using a charged matter field, but rather using thermodynamic work. Assume the availability of a global energy function $U:P\to\RR$ satisfying $dU=\xi+\omega$. A shift in geometric entropy along $\gamma$ shifts the energy $U$ by $U_1 = U(\phi_0,S_1)\mapsto U_2 = U(\phi_0,S_2)$, where $\phi=\phi_0$ is the common initial and final angle. So, since $\gamma$ is an adiabat,
\begin{equation}\label{eq:dU}
  U_2 - U_1 = \int_\gamma(\xi + \omega) = \int_\gamma\omega.
\end{equation}
Thus, the change in energy due to geometric entropy production is associated with a measurable change in the quantity of work done by the process.

An explicit example can be constructed for a simple gas coupled to a rotor with work configuration space given by $M=S^1$, such as a Wankel engine. Let $\alpha=\xi/T$ be a principal connection, with the smooth function $T:P\to\RR$ defined in local coordinates $(\phi,S)$ by $T(\phi,S) = T(S) = T_0\exp(S/C)$. Let $\alpha$ be defined by
\begin{equation}
  \alpha := dS - \kappa d\phi
\end{equation}
for some $\kappa>0$. Then $\xi = T\alpha = TdS - T\kappa d\phi$. Define energy and work respectively by
\begin{equation}\label{def:U-and-w}
  U := CT, \qquad \omega := \kappa Td\phi. 
\end{equation}
Then the First Law is satisfied, $dU = CdT = TdS = \xi + \omega$, where the second equality applies the chain rule to the definition of $T$. However, instead of undergoing an ordinary pressure-volume cycle, the work $\omega=\kappa Td\phi$ corresponds to an external application of torque $\tau=\kappa T$ to the rotor, which will be the source of the change in the system's energy.

Now consider an adiabat $\gamma:[0,1]\to P$ such that $\lambda=\pi\circ\gamma$ is closed. Since $\beta:=dS -\alpha = \kappa d\phi$, geometric entropy is given as in Equation \eqref{eq:geometric-entropy-production} by
\begin{equation}
  \mathcal{E}_\lambda = \int_\lambda B = \int_\gamma \beta = \int_\gamma \kappa d\phi = 2\pi\kappa >0,
\end{equation}
where we assume a winding number of one. Using the definition of $T$, the relationship between the initial and final temperatures is $T_2 = T_1 \exp(2\pi\kappa/C)$. The work associated with this process is then given by Equation \eqref{eq:dU} and the definition of $U$ in Equation \eqref{def:U-and-w},
\begin{equation}
  \int_\gamma\omega = U_2 - U_1 = C(T_2 - T_1).
\end{equation}
Suppose the gas consists of half a litre of air at $1$atm and room temperature, and thus set $C=5nR/2\approx 0.4J/K$ (using the diatomic gas approximation) and $T_1=300K$. A small applied torque $\tau=\kappa T$ with $\kappa = 10^{-3}J/K$ then gives $T_2=304K$, so the process results in a change of temperature of about four degrees, with total work given by $W=C\Delta T\approx1.9J$.

Geometric entropy accumulates locally, analogous to the local accumulation of geometric phase when the Aharonov-Bohm effect is expressed in the unitary gauge.\cite{wallace2014deflating,wallace2024gf} The work done by powering the rotor through its cycle gives rise to an incremental change in temperature. This shifts the system vertically through the entropy coordinate, in a neighbourhood of every point, in spite of the fact that there is no heat exchange. What is particularly unusual is that the system is in local equilibrium in the sense that $\mathcal{F}=0$ and $\xi=TdS_B$ for some local entropy function $S_B$, in a neighbourhood $B$ of every point. Nevertheless, there is no \emph{global} entropy function $S$ satisfying $\xi=TdS$. The vertical fibre coordinate function $S$ still provides a well-defined notion of entropy, and a change in entropy along those fibres is still given by the gauge invariant physical observable $\mathcal{E}_\lambda$. However, the coordinate function is not related to heat globally by $\xi=TdS$. In this sense the system can be viewed as out of `global' equilibrium. 

\section{Further directions}

The gauge theoretic structure of thermodynamics arises out of a single physical distinction, between observable and unobservable contributions to energy: first in terms of a fibre bundle, and second in terms of the concepts of heat and work. The definition of heat in this picture guarantees that it defines an Ehresmann connection, and gives rise to a principal $\RR$-bundle with a principal connection, given the assumption that adiabats remain adiabatic under shifts in `hidden' energy along the fibres. Local entropy and temperature functions appear precisely in the local equilibrium case of vanishing curvature and no change in entropy around local adiabatic work cycles, with negative curvature giving rise to the Second Law. Non-trivial cohomology can give rise to an analogue of geometric phase.

The expression of the law of entropy non-decreasing entropy could potentially be explored in ordinary Yang--Mills gauge theories like electromagnetism. The expression of phase produced by magnetic flux through a loop has a very similar mathematical character to our construction of the Second Law. In the other direction, geometric phase has recently gained interest in the theory of thermal transport, both in theoretical and experimental applications.\cite{RenHanggiLi2010,wang2022geometric} It would be of interest to connect geometric phase in these discussions to the geometric entropy of thermodynamics as a gauge theory.

A number of further opportunities are opened up by this perspective. First, natural links can be drawn between the gauge framework and existing approaches to the geometry of thermodynamics. For example, the order-theoretic framework of \citeauthor{liebyngvason1998n}\cite{liebyngvason1998n,LiebYngvason1999} appears to provide the essential structure needed to view thermodynamics as a gauge theory in the equilibrium case. For simple systems, they describe equilibrium state space as coordinatised by the variables $(U,V_1,\dots,V_m)$. Viewing that space as a manifold $P$, the projection $\pi:(U,V_1,\dots,V_m)\to(V_1,\dots,V_m)$ onto the manifold of work coordinates $M$ can naturally be assumed to be a fibre bundle. Their axiom (S2)\cite{LiebYngvason1999} provides a hyperplane at every point of $P$ for the `forward adiabatic' sector of their adiabatic accessibility relation, analogous to the horizontal distribution given by an Ehresmann connection $H=\ker\xi$. One could explore further relationships between the two frameworks from this perspective.

Similar links can be drawn between the gauge framework and the contact geometry framework for thermodynamics. On the latter approach, the manifold $P$ of equilibrium states is an $n$-dimensional Legendrian submanifold, with local coordinates $(S,V_1,\dots,V_m)$ and $m=n-1$, on which the contact form $\Theta := dU - TdS  + \sum_{i=1}^{m}P_idV_i$ vanishes. Defining $\xi:=TdS$ and $\omega:=-\sum_iP_idV_i$, the manifold $P$ is thus a surface on which $dU = \xi + \omega$ holds. Locally one can then take the projection $\pi:P\to M$ onto the manifold $M$ of $V$ configurations to be a fibre bundle. Choosing the action of $G=(\RR,+)$ to be $R_r(S,V)=(S+r,V)$, it follows that if $T$ is non-vanishing, then $\alpha:=\xi/T$ is a principal connection.

A final topic of foundational interest is the relationship between thermodynamics and classical and quantum statistical mechanics. The gauge theoretic formulation of thermodynamics provides a precise common framework for the study of that relationship, both in equilibrium and non-equilibrium contexts. This could provide new opportunities to study the statistical origin of thermodynamics using tools from gauge physics, including through the use of field equations.


\section*{Data Availability Statement}

Data sharing is not applicable to this article as no new data were created or analyzed in this study.

\section*{References}

%

\end{document}